\documentclass[12pt]{article}
\usepackage{fullpage,setspace,boxit,graphicx}
\begin{document}
\begin{center}
{\Large\bf Brownian Motion for the School-Going Child}\\
\bigskip\bigskip
{\large R.S. Bhalerao \\
{\it Tata Institute of Fundamental Research, Mumbai, India}
}
\end{center}

\bigskip
{\noindent{\bf I. Introduction}}
\bigskip

Let us do a ``thought experiment''. What is a thought
experiment? It is an experiment carried out in thought only. It may or
may not be feasible in practice, but by imagining it one hopes to
learn something useful. There is a German word for it which is
commonly used: {\it Gedankenexperiment}. It was perhaps A. Einstein who
popularized this word by his many gedankenexperiments in the theory of
relativity and in quantum mechanics.

Coming back to our thought experiment: Imagine a dark, cloudy,
moonless night, and suppose there is a power failure in the entire
city. You are sitting in your fourth floor apartment thinking and
worrying about your physics test tomorrow. Suddenly you hear a
commotion downstairs. You somehow manage to find your torch and rush
to the window. Now suppose your torch acts funny: it turns on only for
a moment, every 15 seconds. Initially, i.e., at time $t=0$ seconds,
you see a man standing in the large open space in front of your
building. Before you make out what is happening, your torch is off.
Next time it lights up, i.e., at $t=15$ sec, you see him at a slightly
different location. At $t=30$ sec, he is somewhere else and has
changed his direction too. At $t=45$ sec, he has again changed his
location and direction. You have no idea what is going on. But, you
continue observing him for some more time. When the lights come back,
you mark his positions on a piece of paper (see Fig. 1). At $t=0$, he
is at point A, at $t=15$, he is at B, at $t=30$, he is at C, and so
on. Connect point A to B, B to C, C to D, and so on, by straight
lines. (Go ahead, grab a pencil and do it.) What do you see? A zigzag
path.

\vspace {-1.84cm}
\begin{center}
\begin{figure}[hb!]
\centerline{\includegraphics*{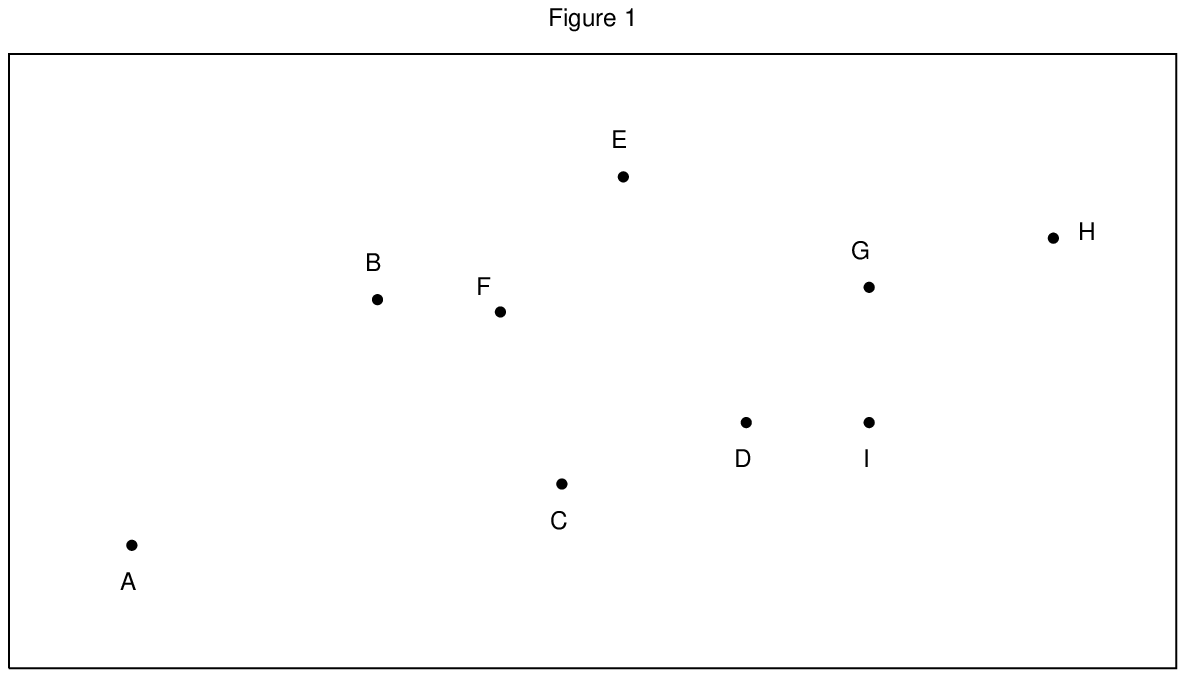}}
\end{figure}
\end{center}

\newpage
What do you think was going on? Did you say ``a drunken man
wandering around the square''? Right. That was easy. One does not need
an Einstein's IQ to figure that out. In physicists' language, the
above is an example of a {\it random walk in two dimensions}: {\bf
two} dimensions because the open area in front of your building has a
length and a breadth. (Strictly speaking, a walk can be said to be
random if the direction of each step is completely independent of the
preceding step. For simplicity, the steps may be taken to be of equal
length.) Before you read further, close your eyes and imagine a
random walk in 1 dimension and then a random walk in 3 dimensions.

\bigskip
\begin{boxit}
\medskip
{\bf Random walk in one dimension:}

Here is an experiment you can do yourselves. You will need a plain
paper, a ruler, a pencil, a one-rupee coin, a small stone and a lot of
patience. Draw a number line with markings at $-10, -9, ..., ~0, ...,
~9, 10$. Place the stone at $0$. Toss the coin. The rule is, if it is
heads (H), the stone is moved one place to the right and if it is
tails (T), it is moved one place to the left. For example, if you get
H, H, T, ..., the stone moves from $0$ to $1$, $1$ to $2$, $2$ to $1$,
... . Toss the coin 10 times. Note the final position of the
stone. Call it $x_1$. Obviously, $-10 \leq x_1 \leq 10$.

Replace the stone at $0$ and repeat the experiment. Again note the
final position of the stone. Call it $x_2$. Obviously, $x_2$ may or
may not be equal to $x_1$.

If you were to repeat this experiment a very large number of times,
say 1000 times, and then take the average ($\bar x$) of $x_1, ~x_2,
~x_3, ~..., ~x_{1000}$, what result do you think you will get? Since
each toss of the coin is equally likely to result in a H or a T, $x_1,
~x_2, ~x_3, ~..., ~x_{1000}$ will be distributed symmetrically around
the origin $0$. Hence $\bar x$ is most likely to be zero.

Interestingly, however, the average ($\overline {x^2}$) of $x_1^2,
~x_2^2, ~x_3^2, ~..., ~x_{1000}^2$, will not be zero, since these are
all non-negative numbers. In fact, $\overline {x^2}$ turns out to be
equal to the number of times you toss the coin in each experiment,
which is also equal to the number of steps ($N$) in the random
walk. (This is 10 in our experiment.) Thus $\overline{x^2}=N$ or
$\left( \overline{x^2} \right)^{1/2}=N^{1/2}$. Since the
left-hand-side is the square root of the mean (= average) of the
squares, it is called the {\it rms} displacement and is denoted by
$x_{\rm rms}$. Thus $x_{\rm rms}=N^{1/2}$.

What is the meaning of the statement $\bar x=0$, but $x_{\rm
rms}=N^{1/2}$? It means, in a random walk, the object is as
likely to be found on one side of the starting point as on the other,
making $\bar x$ vanish. But at the same time, as the number of
steps increases, the object is likely to be found farther and farther
from the starting point.

Equivalently, imagine 1000 drunkards standing at the origins of 1000
parallel lines, and then starting simultaneously their random walks
along these lines. If you observe them after a while, there will be
nearly as many of them to the right of the centres as there are to the
left. Moreover, the longer you observe them, the farther they are
likely to drift from the centre. 

{\bf Conclusions:} (a) $\bar x = 0$. (b) $x_{\rm rms}=N^{1/2}$ if each
step is of unit length. (c) $x_{\rm rms}=N^{1/2}l$ if each step is of
length $l$.

\bigskip
\end{boxit}

Let us perform another thought experiment. Suppose you are sitting in
a big stadium, watching a game of football or hockey, being played
between two equally good teams. As in the previous thought experiment,
you mark on a piece of paper the position of the ball every 15
seconds, and then connect these positions in sequence. What do you
see? Again a zigzag path. The ball is moving almost like the drunken
man. Would you say the ball is drunk? Of course, not. The ball is
moving that way because it is being hit repeatedly by the players in
two competing teams. This is another example of an (almost) random
motion in two dimensions.

Want to impress someone? Remember this: Random processes are also
called stochastic processes. Chance or probability plays an
essential role in these processes.

What you learnt above is the ABC of the branch of physics,
called {\it Statistical Mechanics}.

\bigskip
{\noindent{\bf II. History}}
\bigskip

{\small\it
He is happiest who hath power to gather wisdom from a flower
--- Mary Howitt (1799 - 1888).
}
\smallskip

Now I want to describe a real (not a gedanken) experiment. Robert
Brown was a British botanist. In 1827, he observed through a
microscope pollen grains of some flowering plants. To his surprise, he
noticed that tiny particles suspended within the fluid of pollen
grains were moving in a haphazard fashion.$^1$ If you were Robert
Brown, how would you understand this observation? (Remember, science
in 1827 was not as advanced as it is today. Many things written in
your science textbook were not known then.) Would you suspect that the
pollen grain is alive? Or would you get excited at the thought that
you have discovered the very essence of life or a latent life force within
every pollen? Or perhaps this is just another property of organic
matter? What other experiments would you perform to test your
suspicions?

Brown repeated his experiment with other fine particles including the
dust of igneous rocks, which is as inorganic as could be. He found
that any fine particle suspended in water executes a similar random
motion. This phenomenon is now called {\it Brownian Motion}. Figure 2
shows the result of an actual experiment: the positions of the
particle were recorded at intervals of 30 seconds. (From J. Perrin,
{\it Atoms}, D. Van Nostrand Co., Inc., 1923.) Similar observations
were made for tiny particles suspended in {\it gases}.

Scientists in the 19th century were puzzled by this mysterious
phenomenon. They attempted to understand it with the help of ideas
such as convection currents, evaporation, interaction with 
incident light,
electrical forces, etc. But they had no satisfactory explanation for
it. With your knowledge of modern science, can you provide a
rudimentary explanation? Obviously, the suspended particle is not
moving on its own unlike the drunkard in our first
gedankenexperiment. Why then is it moving? And why in an erratic way
(see Fig. 2)? Think, before you read further.

\vfill
Want a hint? Recall our second gedankenexperiment.

\begin{center}
\begin{figure}[ht!]
\centerline{\includegraphics*{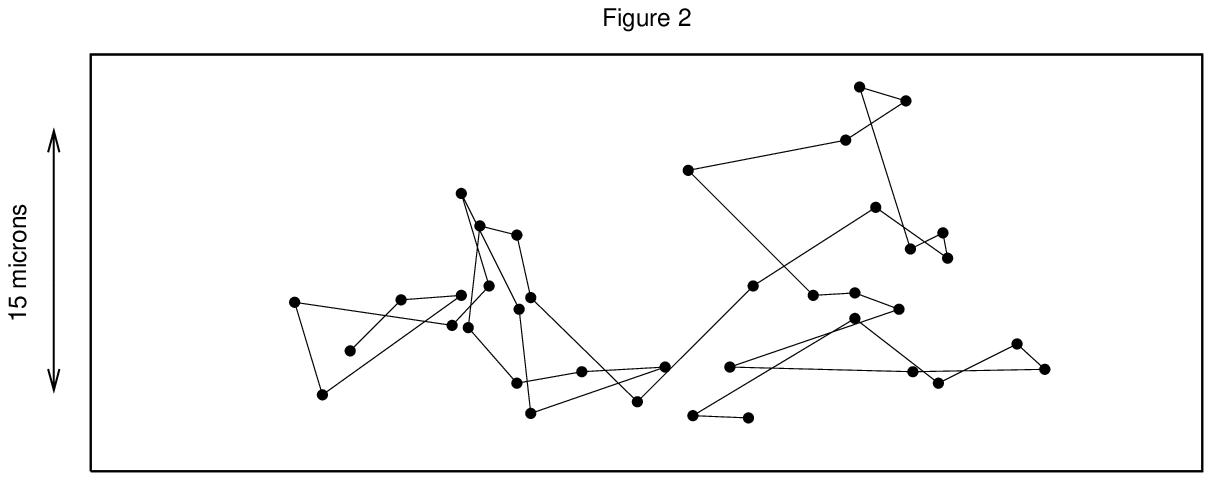}}
\end{figure}
\end{center}

\vspace{-1cm}
{\noindent{\bf III. Basic Understanding}}
\bigskip

If you have not already guessed, here is the rational explanation for
the mysterious jerky movement of tiny particles suspended in fluids,
which made Mr. Brown famous:


$\bullet$ The size --- the radius or diameter --- of the suspended
particle is roughly of the order of a few microns (1 micron =
$10^{-6}$ m). The size of an atom is of the order of $10^{-10}$ m. The
size of a water molecule (H$_2$O) is somewhat larger. Thus the
suspended particle is a monster, about 10000 times bigger compared
to a water molecule. Also note that a spoonful of water contains about
$10^{23}$ water molecules. (The atomic or molecular theory of matter
which says that matter consists of atoms and molecules, is
well-established today. It was not so in 1827!)
 
$\bullet$ You also know that molecules of water (or molecules in any
sample of a liquid or gas) are not at rest. They are perpetually
moving in different directions, some faster than others. As they move,
they keep colliding with each other, which can possibly change their
speeds and directions of motion.

$\bullet$ Now you can very well imagine the fate of the particle
unfortunate enough to be placed in the mad crowd of water
molecules. The poor fellow is getting hit, at any instant, from all
sides, by millions of water molecules. The precise number of water
molecules which hit the particle at any instant, the exact points
where they hit it, their speeds and directions --- all keep changing
from time to time. (It is practically impossible and also 
{\it unnecessary}
to have this information.) This results in a net force which keeps
fluctuating in time, i.e., its magnitude and direction keep changing
from one instant to another. The particle keeps getting kicks in
the direction of the instantaneous net force. The end result is that
its position keeps changing randomly as in Fig. 2.

{\noindent{\bf IV. A Quiz}}
\bigskip

Ready for a quiz? Here are a few easy questions:

(1) Imagine the game of football played by invisible players. (Nothing
except the ball is visible.)

(2) See Fig. 2. If the positions of the particle were recorded every
60 seconds, instead of every 30 seconds, how will the pattern look
like?

(3) How will the Brownian Motion be affected if (a) water is cooled,
or (b) instead of water a more viscous liquid is taken, or (c) the
experiment is done with a bigger particle?

\bigskip
{\noindent{\bf V. Einstein's Contribution}}
\bigskip

You now have a {\it qualitative} understanding of the
Brownian Motion. But that is usually not enough. Scientists like to
develop a {\it quantitative} understanding of a phenomenon. This
allows them to make precise numerical predictions which can be tested
in the laboratory.$^2$ For example, one would like to know how far (on
an average) will the particle drift from its initial position in say 10
minutes? How will its motion be affected if the water is cooled by say
5 C, or if the viscosity of the liquid is increased by 10\%, or if the
particle size is exactly doubled?

In 1905, Einstein published a detailed mathematical theory of the
Brownian Motion, which allowed one to answer these and many other
interesting questions. How did he do it? I will only give you a
flavour of what is involved.

\medskip
\begin{boxit}
\smallskip
{\bf What is an ensemble?:}

Recall that chance or probability plays an important role in random
processes. Hence, in the Introduction, when we discussed the random
walk (see the box), you were asked to do the experiment 1000 times and
then average the results. If you do it only a few times, $\bar x$ may
not vanish and $\overline{x^2}$ may not equal 10. If you are too lazy
to do the experiment 1000 times, there is a way out: Get hold of 1000
friends of yours, ask each of them to prepare a similar experimental
set-up, and let each of them do the experiment only once. If you then
take the average of the results obtained by them, you will find $\bar
x \approx 0$ and $\overline{x^2} \approx 10$.

Similarly, if you observe the Brownian Motion of a particle only a
few times, based on these observations,
you would not be able to make quantitative statements about
its average behaviour. You need to repeat
the experiment a large number of times and take the average of all the
results. Alternately, you could prepare a large assembly of
identical particles, observe each of them once under identical
experimental conditions, and then take the average.

Physicists use the word {\it ensemble} to describe such an imaginary
assembly of a very large number of similarly prepared physical
systems. The average taken over all the members of the ensemble is
called an ensemble average. In the following, when we talk about an
average behaviour of a Brownian particle, we mean an ensemble average.
\medskip
\end{boxit}

\newpage
Let us ask ourselves a few simple questions about the average
behaviour of a tiny particle suspended in a liquid. Taking the
initial location of the particle as the origin, imagine drawing $x$,
$y$ and $z$ axes in the liquid. Let $(x,y,z)$ denote the coordinates
of the particle.

\fbox{$\bar x$}~: 
Where will the particle be after some time? In other words,
what will be the values of $\bar x$, $\bar y$ and $\bar z$ after some
time? (Remember the overhead lines denote ensemble averages.) Since
this is a case of a random walk$^3$ in 3 dimensions, $\bar x=\bar y=
\bar z=0$.

\fbox{$\bar v_x$}~: 
What will be the average velocity of the particle parallel
to the $x$ axis? When we talk of the velocity of a particle, we have
{\it two} things in our mind: its speed (fast or slow) and its
direction of motion. Since the particle is as likely to move in the
positive-$x$ direction as in the negative-$x$ direction, $v_x$ is as
likely to be positive as negative. Hence $\bar v_x=0$.
Similarly, $\bar v_y=\bar v_z=0$.

\fbox{$\overline{v_x^2}$}~:
What will be the value of $\overline{v_x^2}$? This ensemble
average will not be zero since $v_x^2$ is either positive or zero ---
never negative. I already said that molecules of water are not at
rest. They are perpetually moving in different directions, some faster
than others. Now, heat is a form of energy. When we heat water, we
give energy to it. As a result, the water molecules start moving
faster. Their average kinetic energy rises. On the other hand, when we
heat water, its temperature also rises. Thus temperature is a measure
of the average kinetic energy of the water molecules. It turns out
that when the suspended particle is in thermal
equilibrium with the water, its average kinetic energy is proportional
to the temperature:
$$\frac{1}{2}m\overline{v_x^2}=\frac{1}{2}kT,$$ where $m$ is the mass
of the suspended particle, $k$ is a constant and $T$ is the absolute
temperature of the water. Hence $\overline{v_x^2}=kT/m$. This
implies that heavier
particles will have smaller $\overline{v_x^2}$.  Similar statements
can be made about the motion in $y$ and $z$ directions.$^4$

\fbox{$\overline{x^2}$}~: 
How far from the origin will the particle be after some
time? Equivalently, what will be the value of $\overline{x^2}$? Before
I answer this question, note the important complication in the present
problem. Now, not only the direction but also the size of each step is
a variable and is completely independent of the preceding step. (Why?
Remember the fluctuating force mentioned in section III.)

Using the ideas from Statistical Mechanics, Einstein derived the 
following result:
$$\overline {x^2}= \frac{kT}{3 \pi \eta a}t,$$ where $\eta$ is the
viscosity of the liquid, $a$ is the radius of the suspended particle
(assumed to be spherical) and $t$ is the elapsed time. Thus the mean
square displacement $\overline {x^2}$ increases {\it linearly}
with time (i.e., the power of $t$ in the above equation is unity).

Looking at the last equation, can you now answer the question no. (3)
in the Quiz above? Please find out yourselves, before you read the
answers given here:
  
(a) The Brownian Motion is less vigorous in cold water than in hot
water. (b) The Brownian Motion will be damped if water is replaced by
a more viscous liquid. (c) A bigger particle will drift less than a
smaller particle --- we do not notice the Brownian Motion of fish, 
people or boats. Do we?

Using Einstein's result, one can also answer quantitatively the more
specific questions listed at the beginning of section V.

\bigskip
{\noindent{\bf VI. Importance}}
\bigskip

$\bullet$ In 1908, the French physicist Jean-Baptiste Perrin 
verified Einstein's result
experimentally. He measured $\overline {x^2}$ as a function of time
$t$. Knowing the temperature $T$ and viscosity $\eta$ of the
water, and radius $a$ of the particle, he could obtain the value of
the constant $k$. Using this, he obtained a reasonably good value for
Avogadro's number (no. of molecules in a mole of a substance).

$\bullet$ Einstein's explanation of the Brownian Motion and its
subsequent
experimental verification by Perrin$^5$
were historically important because they
provided a convincing evidence for the molecular theory of matter. In
other words, they showed that atoms and molecules are real physical
objects. Skeptics who doubted their existence were silenced.

$\bullet$ Fluctuating force on a Brownian particle is but one example
of a fluctuating physical quantity even when the system is in
equilibrium. Another example is a fluctuating current in some electric
circuits. Einstein's work on the Brownian Motion laid the foundations
of the study of fluctuation phenomena as a branch of statistical
mechanics.

\bigskip\bigskip
We have reached the end of our story of the Brownian Motion.
You must have realized how a lowly pollen grain 
can tell us so much about the constitution of matter. 
Note that nothing
of this would have been possible without the inquisitive mind of the
scientist. The following quotation comes to my mind:

{\it There is something fascinating about science. One gets such
wholesale returns of conjecture out of such a trifling investment of
fact} --- Mark Twain (1835 - 1910).


\newpage
{\small
{\noindent Footnotes:}

$^1$ Sometimes it is wrongly stated that Brown observed irregular
motion of the pollen grains themselves. Secondly, he was not the first
to notice this phenomenon. But he was the first to stress its
ubiquitousness and to rule out its explanations based on the so-called
life force. As a result of his work, this subject was removed from the
realm of biology into the realm of physics.

$^2$ However good a theory may appear, if its predictions do not agree
with experimental data, it is discarded.}

$^3$ Here we assumed that successive time intervals are very small
compared with the observation time but still large enough that the
motion of the suspended particle in any time interval can be
considered to be completely independent of its motion in the preceding
time interval.
 
$^4$ Einstein showed that it is practically impossible to measure
$\overline{v_x^2}$ and suggested that experimentalists should rather
measure $\overline{x^2}$.

$^5$ Perrin was honoured with the Nobel Prize for Physics in 1926,
for this work.

\end{document}